\documentclass[a4paper,11pt]{article}
\pdfoutput=1
\usepackage{jcappub}
\usepackage[T1]{fontenc}
\usepackage{graphicx}
\usepackage{epsfig}

\def\be{\begin{equation}}
\def\ee{\end{equation}}
\newcommand{\bq}{\begin{eqnarray}}
\newcommand{\eq}{\end{eqnarray}}
\newcommand{\bes}{\begin{subequations}}
\newcommand{\ees}{\end{subequations}}

\def\ben{\begin{eqnarray}}
\def\een{\end{eqnarray}}
\def\ba{\begin{array}}
\def\ea{\end{array}}

\newcommand\lsim{\mathrel{\rlap{\lower4pt\hbox{\hskip1pt$\sim$}}
    \raise1pt\hbox{$<$}}}
\newcommand\gsim{\mathrel{\rlap{\lower4pt\hbox{\hskip1pt$\sim$}}
    \raise1pt\hbox{$>$}}}
\newcommand\esim{\mathrel{\rlap{\raise2pt\hbox{\hskip0pt$\sim$}}
    \lower1pt\hbox{$-$}}}

\title{\boldmath Could the dynamics of the Universe be influenced by what is going on inside black holes?}

\author[a,b,c]{P. P. Avelino}

\affiliation[a]{Instituto de Astrof\'{\i}sica e Ci\^encias do Espa{\c c}o, Universidade do Porto, CAUP, Rua das Estrelas, PT4150-762 Porto, Portugal}
\affiliation[b]{Centro de Astrof\'{\i}sica da Universidade do Porto, Rua das Estrelas, PT4150-762 Porto, Portugal}
\affiliation[c]{Departamento de F\'{\i}sica e Astronomia, Faculdade de Ci\^encias, Universidade do Porto, Rua do Campo Alegre 687, PT4169-007 Porto, Portugal}

\emailAdd{pedro.avelino@astro.up.pt}

\abstract{We investigate the potential impact of mass inflation inside black holes on the dynamics of the Universe, considering a recent reformulation of general relativity, proposed in \cite{Kaloper:2013zca}, which prevents the vacuum energy from acting as a gravitational source. The interior dynamics of accreting black holes is studied, at the classical level, using the homogeneous approximation and taking charge as a surrogate for angular momentum. We show that, depending on the accreting fluid properties, mass inflation inside black holes could influence the value of the cosmological constant and thus the dynamics of the Universe. A full assessment of the cosmological role played by black holes will require a deeper understanding of the extremely energetic regimes expected inside real astrophysical black holes, including their relation with the physics of the very early Universe, and may eventually lead to an entirely new paradigm for the origin and evolution of the Universe.}

\begin{document}
\maketitle
\flushbottom

\section{\label{intr}Introduction}

Whether or not a constant energy density should be a source for the gravitational field is still a matter of debate. This fundamental question has been the main motivation for several modified gravity models proposed in the literature \cite{Tseytlin:1990hn,Padmanabhan:2007xy,Kaloper:2013zca,Gabadadze:2014rwa}. Recently, in \cite{Kaloper:2013zca}, a reformulation of general relativity was proposed which prevents the vacuum energy from acting as a gravitational source. In this scenario the value of the cosmological constant depends on a specific space-time average of the trace of the energy-momentum tensor and the universe is predicted to be finite in space and time, the present acceleration being a transient stage before the big crunch \cite{Kaloper:2013zca,Kaloper:2014dqa,Kaloper:2014fca,Avelino:2014nqa}.

In \cite{Kaloper:2014dqa} the authors investigated the possible impact of black holes on the value of the cosmological constant, concluding that it should be very small. In their derivation the authors considered black holes to be described by the Schwarzschild metric, not accounting for the rotation and accretion expected in the case of real astrophysical black holes. In the present paper, we study the possible contribution of black holes to the value of the cosmological constant, considering the more realistic case of accreting black holes and taking charge as a surrogate for angular momentum. We shall focus on the role played by mass inflation, a dramatic phenomenon which is predicted to occur near the inner horizon as a consequence of relativistic counter-streaming between ingoing and outgoing streams \cite{Poisson:1989zz,Ori:1991zz} (see also \cite{Hamilton:2008zz} for a detailed exposition of the physical cause and consequence of mass inflation).

Throughout this paper we use units such that $G=c=4\pi \epsilon_0=1$, where $G$ is the gravitational constant, $c$ is the value of the speed of light in vacuum and $\epsilon_0$ is the vacuum permittivity. We adopt the metric signature $(-,+,+,+)$.

\section{Vacuum energy sequestering and Schwarzschild black holes}

Here we shall consider the action defined in \cite{Kaloper:2013zca} which yields the following equation for the gravitational field
\be
G^{\mu\nu}=8\pi T^{\mu \nu} -  \Lambda g^{\mu\nu}\,,
\ee
where $G^{\mu\nu} \equiv R^{\mu\nu}-g^{\mu\nu}R/2$ are the components of the Einstein tensor, $g^{\mu\nu}$ are the components of the metric, $R^{\mu\nu}$  are the components of the Ricci curvature tensor, $R \equiv {R^{\mu}}_{\nu}$ is the Ricci scalar curvature, $T^{\mu \nu}$ are the components of the energy momentum tensor and the cosmological constant $\Lambda$ is given by
\be
\frac{\Lambda}{2\pi}=\langle {T^{\mu}}_{\mu} \rangle \equiv \frac{\int d^4 x {\sqrt {-g}}{T^{\mu}}_{\mu}}{\int d^4 x {\sqrt {-g}}}\,,
\label{lambda}
\ee
where $g={\rm det}(g_{\mu\nu})$ is the metric determinant. These equations of motion for the gravitational field are invariant under the transformation $T^{\mu \nu} \to T^{\mu \nu}+C g^{\mu\nu}$ where $C$ is an arbitrary real constant. Consequently, any bulk constant energy density is effectively gauged away.

In  \cite{Kaloper:2014dqa} the authors have argued that astrophysical black holes provide a negligible contribution to the value of $\Lambda$. They considered Schwarzschild black holes with a compactified time direction (where time is limited by the total lifetime of the universe) and line element
\be
ds^2=-\left(1-\frac{r_s}{r}\right) dt^2+\frac{dr^2}{1-\frac{r_s}{r}} +r^2 \left(d\theta^2 +\sin^2 \theta d \phi^2  \right)\,,
\ee
where $r_s=2 M$ is the Schwarzschild radius. The interior ($I$) spacetime volume can then be estimated as
\be
\int _{I} d^4 x \sqrt{-g} = 4 \pi \int_0^{\Delta t} dt \int_0^{r_s} dr  r^2  \lsim \frac{4 \pi }{3} r_s^3 t_U \,,
\ee
where $\Delta t$ and $t_U$ are the black hole and Universe lifetimes, respectively.
The  total contribution of the black holes' interior spacetime volume is given by
\be
\sum_{i=1}^{N_{BH}} \int _{I_i} d^4 x \sqrt{-g} = \frac{4 \pi}{3} \sum_{i=1}^{N_{BH}} \Delta t_i r_{si}^3  \lsim \frac{4 \pi }{3} t_U \sum_{i=1}^{N_{BH}} r_{si}^3 \,,
\ee
where $N_{BH}$ is the total number of black holes inside the Hubble volume.
If either the fraction of the total energy in the form of black holes inside the Hubble volume is negligible or the number of black holes inside the Hubble volume is large when the Universe approaches its maximum size then the total contribution of the interior spacetime volume of black holes to $\int {\sqrt {-g}} d^4 x$ is negligible.
On the other hand
\be
\sum_{i=1}^{N_{BH}} \int _{I_i} d^4 x {\sqrt {-g}} {T^{\mu}}_{\mu}   \lsim M_{BH} t_U\,.
\ee
where $M_{BH}$ is the total mass of all black holes. If black-holes provide a negligible contribution to the energy density of the Universe when the Universe approaches its maximum size then the total contribution of black holes to $\int {\sqrt {-g}} {T^{\mu}}_{\mu} d^4 x$ is very small. This in agreement with the result obtained in \cite{Kaloper:2014dqa} which led the authors to suggest a negligible contribution of black holes to the value of $\Lambda$.  

\section{Rotating vs charged black holes, mass inflation and the value of $\Lambda$}

Since realistic black holes are not isolated and are expected to rotate rapidly, the computation of their internal structure must take into account accretion and angular momentum.  The relativistic counter-streaming between ingoing and outgoing streams inside rotating Kerr black holes has been shown to be responsible for an exponential growth of the Misner-Sharp mass, a process known as mass inflation \cite{Poisson:1989zz,Ori:1991zz}. The interior structure of a charged black hole resembles that of a rotating black hole, with the negative pressure associated to the electric field generating a gravitational repulsion analogous to that produced by the centrifugal force in a rotating black hole. Hence, although the study of generic perturbations around an axisymmetric spacetime is extremely difficult, it is common to model black hole interiors using charge as a surrogate for angular momentum, considering accreting spherically symmetric charged black holes (see, for example, \cite{Hansen:2005am,Avelino:2009vv}). 

The spherically symmetric line element
\be
ds^2 = g_{tt}(r)dt^2+g_{rr}(r)dr^2+r^2 \left(d\theta^2+\sin^2 \theta d\phi^2\right)\,,
\label{hom}
\ee
with
\bq
g_{tt}&=&-\left(1-\frac{2M}{r}+\frac{Q^2}{r^2}\right)\,,\\
g_{rr}&=&-\frac{1}{g_{rr}}\,,
\eq
describes the empty space-time geometry in and around spherically symmetric black holes of mass $M$ and charge $Q$. The outer ($r_+$) and inner ($r_-$) horizons are located at
\be
r_{\pm}=\left(M \pm {\sqrt{M^2-Q^2}}\right)\,.
\ee

The interior structure of black holes is dramatically affected by accretion. Here we shall consider charged black holes  accreting a perfect fluid with energy-momentum tensor
\be
^f{T^{\mu}}_{\nu} = (\rho + p) U^{\mu} U_{\nu} + p {\delta^{\mu}}_{\nu}\,,
\label{pf}
\ee
where $\rho$ and $p$ are the proper density and pressure, $w=p/\rho$ is the equation of state parameter and $U^{\mu}$ are the components of the 4-velocity. Eq. (\ref{pf}) can be used to describe the energy-momentum tensor of a canonical scalar field $\phi$ 
\be
^\phi{T^{\mu}}_{\nu} = \phi^{,\mu} \phi_{,\nu}+(X-V(\phi)){\delta^{\mu}}_{\nu}\,,
\ee
where $X=-\phi^{,\mu} \phi_{,\mu}/2$ and $V(\phi)$ is the scalar field potential. This correspondence can be made explicit by the identifications
\bq
U_\mu&=&\frac{\phi_{,\mu}}{\sqrt {2X}}\,, \rho=X+V(\phi) \,, p=X-V(\phi)\,,
\eq
and it is valid as long as $\phi_{,\mu}$ is timelike.

We shall use the homogeneous approximation in order to compute the black hole's interior structure, thus assuming that all quantities are functions only of the radial (timelike) coordinate $r$ (in which case the line element is still given by Eq. (\ref{hom})). This approximation not only simplifies the mathematics but has also been shown to provide an accurate description of some of the most important aspects of mass inflation. In the homogeneous approximation, a canonical scalar field behaves as a perfect fluid and the energy density $\rho$ may be computed as a function of $r$ and the metric coefficients from energy-momentum conservation (${T^{\mu \nu}}_{;\nu}=0$) as
\be
\rho= - T_{r}^{r}=\rho_i\left(\frac{g_{tti}}{g_{tt}}\right)^{(1+w)/2} \left(\frac{r_i}{r}\right)^{2(1+w)}\,,
\ee
where have taken $U_r=-1$, $U_t=U_\theta=U_\phi=0$, we have assumed a constant $w$ and $\rho_i$, $g_{tti}$ and $r_i$ are integration constants where the subscript $i$ refers to the the values of the various quantities at the surface $r=r_i$.
The non-zero components of the energy-momentum tensor of the electric field corresponding to a constant charge $Q$ are
\be
{^eT^{r}}_{r}={^eT^{t}}_{t}=-{^eT^{\theta}}_{\theta}=-{^eT^{\phi}}_{\phi}=-\frac{Q^2}{8\pi r^4}\,. 
\ee

The $rr$ and $tt$ components of the Einstein equations are given by
\bq
\frac{g_{tt}-g_{rr}g_{tt}+rg_{tt}'}{r^2 g_{rr} g_{tt}}&=&8\pi\left({^fT^{r}}_{r}+ {^eT^{r}}_{r}\right)\,, \label{massinf1}\\
\frac{g_{rr}-{g_{rr}}^2-rg_{rr}'}{r^2 {g_{rr}}^2}&=&8\pi\left({^fT^{t}}_{t}+{^eT^{t}}_{t}\right) \label{massinf2}\,,
\eq
where a prime denotes a derivative with respect to $r$ (here we neglect the very minor role played by a small  cosmological constant inside black holes).

\begin{figure}
\begin{center}
\includegraphics*[width=9cm]{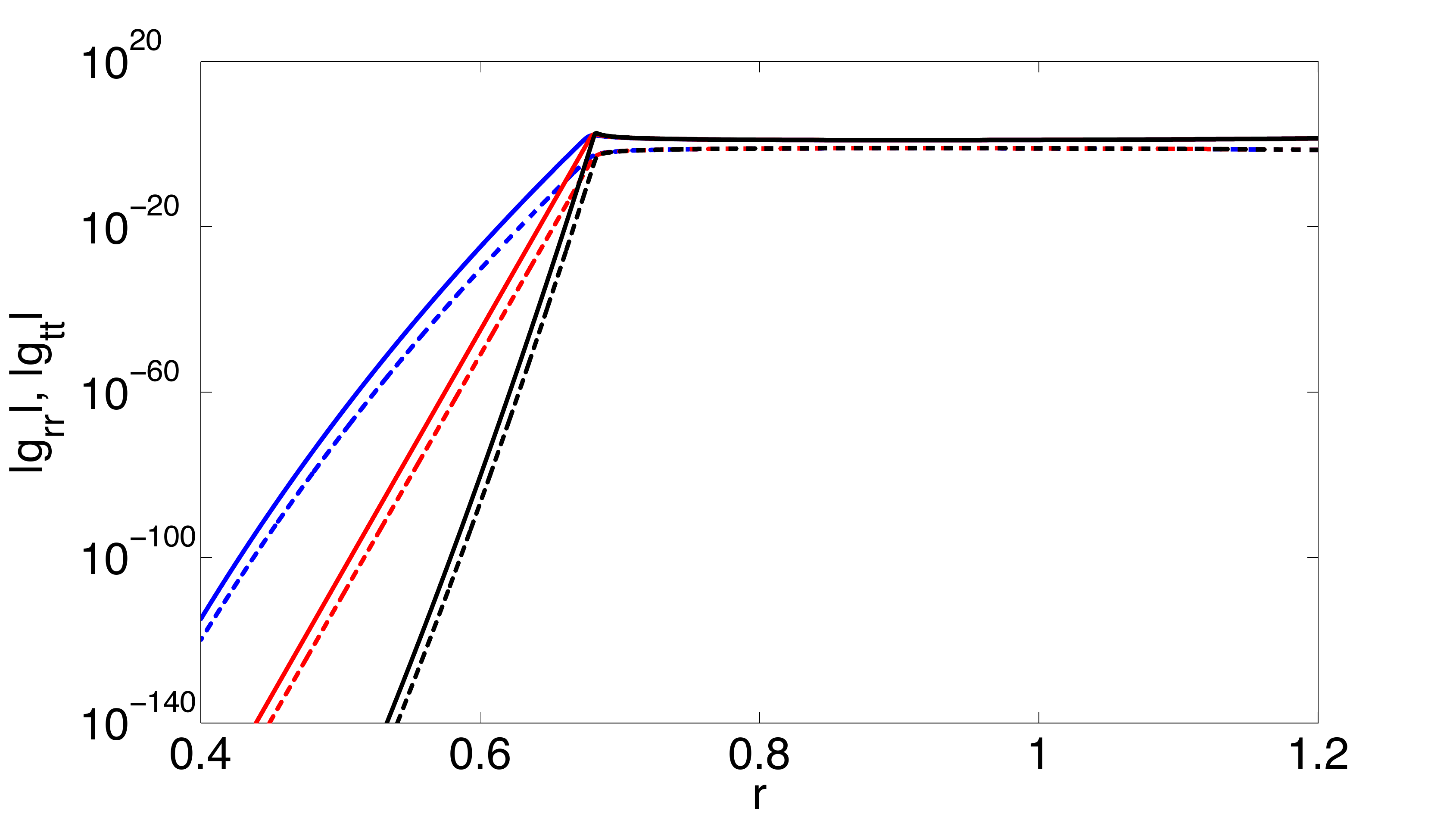}
\end{center}
\caption{\label{fig1}Fig. \ref{fig1} shows the evolution of $|g_{rr}|$, $|g_{tt}|$ with $r$ (solid and dashed lines, respectively) assuming $r_i=0.95 r_-$, $\rho_i=5 \times 10^{-4}$, $g_{tti}=-(1-2M/r_i+Q^2/r_i^2)$, $g_{rri}=-1/g_{tti}$, $M=1$, $Q=0.95$, $w_\parallel=1$ and $w_\perp=0$, $0.5$ and $1$ (black, red and blue lines, from bottom to up, respectively). In all cases the behaviour of $|g_{rr}|$ and $|g_{tt}|$ for $r<r_-$ indicates that mass inflation is taking place.}
\end{figure}

In the following we shall consider the case of a spherically symmetric charged black hole accreting a massless scalar field (taking $V(\phi)=0$, so that $w=1$), a case where mass inflation has been shown to occur (see, for example, \cite{Hansen:2005am,Avelino:2009vv}). Adding Eqs. (\ref{massinf1}) and (\ref{massinf2}) one obtains
\be
\left(r^2\frac{g_{tt}}{g_{rr}}\right)'=2g_{tt}\left(r-\frac{Q^2}{r}\right)\,,
\label{sum}
\ee
assuming that $w=1$.
The value of $g_{tt}$ becomes very small during mass inflation (see Fig. \ref{fig1}) and, consequently, the r.h.s. of Eq. (\ref{sum}) becomes negligible for $r<r_-$, thus leading to the following analytic approximation
\be
{\sqrt {-\frac{g_{rr}}{g_{tt}}}}=A\left(\frac{r}{r_-}\right)\,,
\label{qev}
\ee
where $A>0$ is a constant. Note that, in the mass inflation region, for $r \sim r_-$, $g_{rr}$ and $g_{tt}$ are roughly proportional to each other, as shown in \cite{Avelino:2009vv,Avelino:2011ee}.

A very rough estimate of the constant $A$ can be made by taking into account that mass inflation starts for $r \sim r_-$ when the two terms on the r.h.s. of Eq. (\ref{massinf1}) become of the same order. This happens when $\rho \sim Q^2/(8\pi r^4)$, or equivalently (for $w=1$), when
\be
\frac{1}{g_{tt}} \sim C = \frac{Q^2}{8\pi g_{tti} \rho_i r_i^4}\,.
\ee
Taking into account that at the start of mass inflation $g_{rr} g_{tt} \sim -1$, we find that $A \sim C$.

We may now estimate the contribution of the the interior region with $r \in ]0,r_-[$, of a single black hole, to $\int_I  d^4 x {\sqrt {-g}}$ and $\int_I  d^4 x {\sqrt {-g}} {T^{\mu}}_{\mu}$. The first integral is given by 
\be
\Delta t \int_0^{r_-} {\sqrt {- g_{rr} g_{tt}}}  r^2 dr \ll \Delta t \int_0^{r_-} r^2 dr\,,
\ee
thus making the contribution of the region with $r \in ]0,r_-[$ to $\int_I  d^4 x {\sqrt {-g}}$ very small. On the other hand, 
\bq
\int_{I(]0,r_-[)}  d^4 x {\sqrt {-g}} {T^{\mu}}_{\mu}&=&8\pi\int_0^{\Delta t} dt \int_0^{r_-} {\sqrt {- g_{rr} g_{tt}}} \rho r^2 dr \nonumber \\ 
&=&  \frac{Q^2}{r_-}  \Delta t \int_0^{r_-} \frac{dr}{r} \,.
\eq
which diverges logarithmically at $r=0$ (note that ${T^{\mu}}_{\mu}=2\rho$ if $w=1$), thus suggesting that the interior dynamics of black holes could have a significant impact on the value of $\Lambda$.

Now we shall consider the most general case with symmetrically equal ingoing and outgoing fluxes, for which the homogeneous approximation remains valid. The non-zero components of the most general energy-momentum tensor consistent with spherical symmetry are given by
\be
T_{r}^{r}=-\rho\,,  T_{t}^{t}=p_\parallel \,, T_{\theta}^{\theta}=T_{\phi}^{\phi}=p_\perp\,.
\ee
Energy-momentum conservation implies that
\be
\rho= - T_{r}^{r}=\rho_i\left(\frac{g_{tti}}{g_{tt}}\right)^{(1+w_\parallel)/2} \left(\frac{r_i}{r}\right)^{2(1+w_\perp)}\ , 
\ee
where $w_\parallel=p_\parallel/\rho$ and  $w_\perp=p_\perp/\rho$. Again mass inflation occurs for $w_\parallel=1$, even if $w_\perp \neq 1$. In this case, Eq. (\ref{qev}) remains valid but now
\be
A \sim C = \frac{Q^2}{8 \pi g_{tti} \rho_i r_i^{2(1+w_\perp)} r_-^{2(1-w_\perp)}}\,,
\label{ac}
\ee
and the integral
\bq
\int_{I(]0,r_-[)}  d^4 x {\sqrt {-g}} {T^{\mu}}_{\mu} &\sim& 4 \pi \int_0^{\Delta t} dt \int_0^{r_-} {\sqrt {- g_{rr} g_{tt}}} \rho r^2 dr \nonumber \\
&\sim& \frac{Q^2}{r_-}  w_\perp \Delta t \int_0^{r_-} \frac{dr}{r^{2 w_\perp-1}} \nonumber \\
&\sim& \frac{Q^2}{r_-} \frac{w_\perp}{1-w_\perp}\,,
\eq
no longer diverges for a constant $w_\perp \neq 1$  (note that ${T^{\mu}}_{\mu}=2 w_\perp \rho$ if $w_\parallel=1$).

Fig. \ref{fig1} shows the evolution of $|g_{rr}|$, $|g_{tt}|$ with $r$ (solid and dashed lines, respectively) assuming $r_i=0.95 r_-$, $\rho_i=5 \times 10^{-4}$, $g_{tti}=-(1-2M/r_i+Q^2/r_i^2)$, $g_{rri}=-1/g_{tti}$, $w_\parallel=1$ and $w_\perp=0$, $0.5$ and $1$ (black, red and blue lines, from bottom to up, respectively). Without loss of generality, we choose units such that $M=1$. Observational evidence suggest that real astrophysical black holes may have nearly extremal spins \cite{McClintock:2006xd,Brenneman:2006hw,Miller:2009cw,McClintock:2011zq,Brenneman:2011wz,Gou:2011nq}. Hence, we assume the black hole charge to be $Q=0.95$, close to the extremal value of unity. Fig. \ref{fig1} shows that if $w_\parallel=1$ then mass inflation takes place for any value of $w_\perp$.

\begin{figure}
\begin{center}
\includegraphics*[width=9cm]{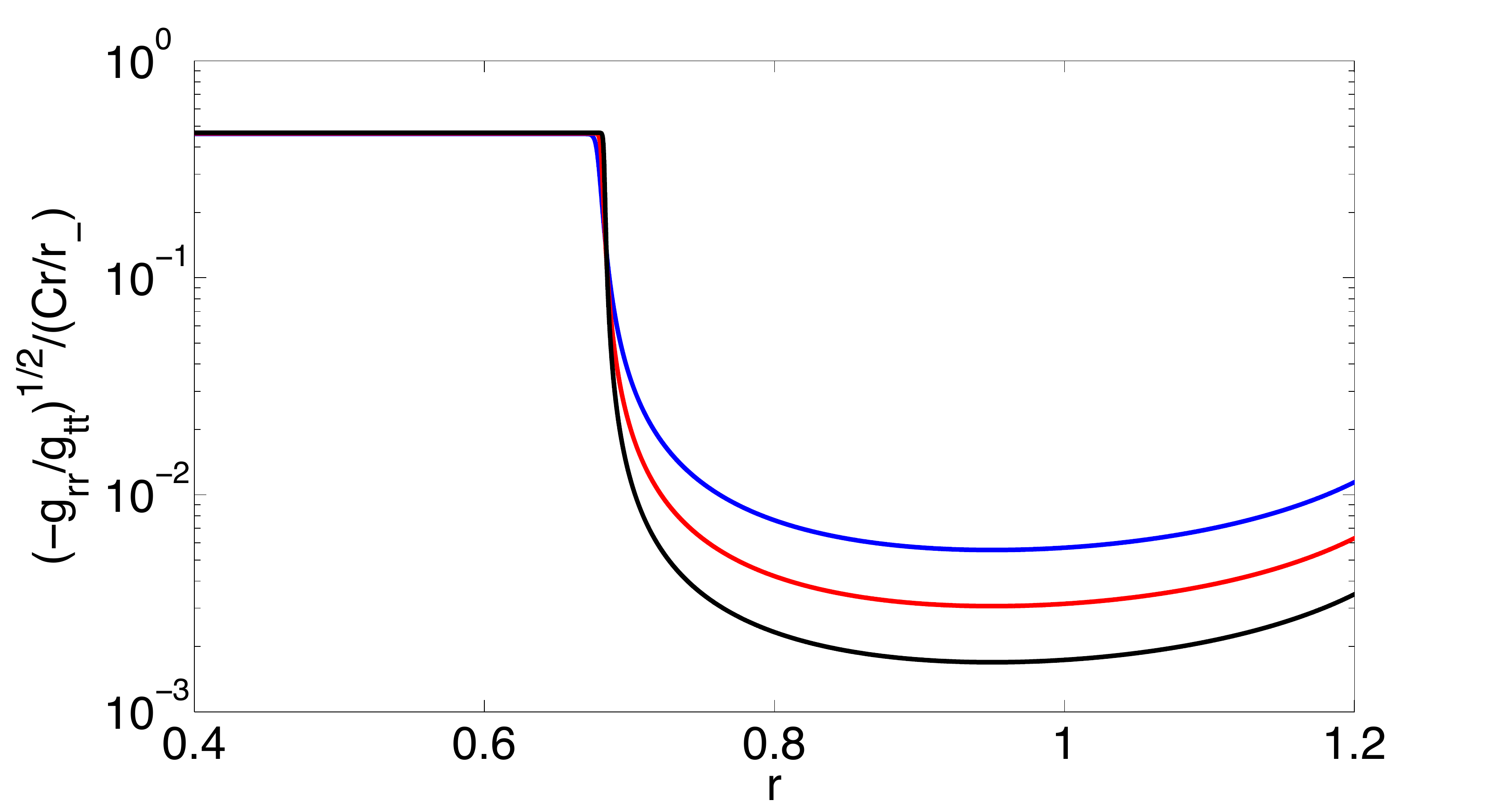}
\end{center}
\caption{\label{fig2} The evolution of ${\sqrt {-g_{rr}/g_{tt}}}/(Cr/r_-)$ with $r$ for the models shown in Fig. \ref{fig1} with $w_\parallel=1$ and $w_\perp=0$, $0.5$ and $1$ (black, red and blue lines, from bottom to up, respectively). As expected from the analytical calculation ${\sqrt {-g_{rr}/g_{tt}}}/(Cr/r_-)$ tends to a constant of order unity when $r$ becomes smaller than $r_-$.}
\end{figure}

Fig. \ref{fig2} shows the evolution of ${\sqrt {-g_{rr}/g_{tt}}}/(Cr/r_-)$ with $r$ for the models shown in Fig. \ref{fig1}. As expected from the analytical calculation, given in Eqs. (\ref{qev}) and (\ref{ac}), ${\sqrt {-g_{rr}/g_{tt}}}/(Cr/r_-)$ tends to a constant of order unity when $r$ becomes smaller than $r_-$.

In the extremal $Q=M$ limit (for which $r_-=r_+=M$)
\be
\frac{1}{M}\int_{I(]0,r_-[)}  d^4 x {\sqrt {-g}} {T^{\mu}}_{\mu} \sim \frac{w_\perp}{2(1-w_\perp)}\,,
\ee
which becomes equal to unity for $w_\perp = 2/3$. Hence, the region with $r \in ]0,r_-[$ dominates the contribution of black holes to the value of $\Lambda$ for $w_\perp > 2/3$.

\section{\label{conc}Conclusions}

We investigated the role played by accreting black holes in the determination of the cosmological constant, considering a recent reformulation of general relativity incorporating a vacuum energy sequestering mechanism.  The interior black hole dynamics was studied, at the classical level, using the homogeneous approximation and taking charge as a surrogate for angular momentum. We have shown that, depending on the specific properties of the accreting fluid, the changes to the internal structure of black holes associated to mass inflation may influence the value of the cosmological constant $\Lambda$ and, consequently, the evolution of the Universe. 

Mass inflation in the vicinity of the inner horizon of charged or rotating Black Holes acts as a particle accelerator, with Planckian and trans-Planckian physics becoming relevant much before the central singularity. Hence, quantum gravity effects are expected to be relevant and should be taken into account in future analysis. The extremely energetic regimes attained during mass inflation are also expected in the very early Universe. This has led to the conjecture of the possible creation of new expanding universes inside Black Holes \cite{Smolin:1994vb,Avelino:2011ee}, a possibility that might change the way we perceive the origin and evolution of the Universe.

%%%%%%%%%%%%%%%%%%%%%%%%%%%%%%%%%%%%%%%%%%%%%%%%%%%%%%%%%%
\acknowledgments

P.P.A. is supported by Funda{\c c}\~ao para a Ci\^encia e a Tecnologia (FCT) through the Investigador FCT contract reference IF/00863/2012 and POPH/FSE (EC) by FEDER funding through the program Programa Operacional de Factores de Competitividade - COMPETE. Funding of this work was also provided by the FCT grant UID/FIS/04434/2013.

%%%%%%%%%%%%%%%%%%%%%%%%%%%%%%%%%%%%%%%%%%%%%%%%%%%%%%%%%%

\bibliographystyle{plain}
\bibliography{BH}

\end{document}